\documentclass[aps,prl,preprint,groupedaddress,superscriptaddress,showpacs]{revtex4-1}

\usepackage{amsmath}
\usepackage{amssymb}
\usepackage{graphicx}
\usepackage{dcolumn}
\usepackage{bm}
\usepackage[usenames]{color}
\usepackage[normalem]{ulem}
\usepackage[none]{hyphenat}

\begin{document}

\title{Magnetic proximity effect at the 3D topological insulator/magnetic insulator interface}

\author{S.V. Eremeev}
 \affiliation{Institute of Strength Physics and Materials Science,
634021, Tomsk, Russia}
 \affiliation{Tomsk State University, 634050 Tomsk, Russia}

\author{V.N. Men'shov}
 \affiliation{NRC Kurchatov Institute, Kurchatov Sqr. 1, 123182 Moscow, Russia}

\author{V.V. Tugushev}
 \affiliation{NRC Kurchatov Institute, Kurchatov Sqr. 1, 123182 Moscow, Russia}
 \affiliation{A.M. Prokhorov General Physics Institute, Vavilov str. 38, 119991
Moscow, Russia}

\author{P.~M. Echenique}
 \affiliation{Donostia International Physics Center (DIPC), 20018 San Sebasti\'an/Donostia, Basque Country, Spain\\}
\affiliation{Departamento de F\'{\i}sica de Materiales UPV/EHU,
Centro de F\'{\i}sica de Materiales CFM - MPC and Centro Mixto
CSIC-UPV/EHU, 20080 San Sebasti\'an/Donostia, Basque Country, Spain}

\author{E.V. Chulkov}
\affiliation{Donostia International Physics Center (DIPC),
             20018 San Sebasti\'an/Donostia, Basque Country,
             Spain\\}
\affiliation{Departamento de F\'{\i}sica de Materiales UPV/EHU,
Centro de F\'{\i}sica de Materiales CFM - MPC and Centro Mixto
CSIC-UPV/EHU, 20080 San Sebasti\'an/Donostia, Basque Country, Spain}

\begin{abstract}
The magnetic proximity effect is a fundamental feature of
heterostructures composed of layers of topological insulators and
magnetic materials since it underlies many potential applications in
devices with novel quantum functionality. Within density functional
theory we study magnetic proximity effect at the 3D topological
insulator/magnetic insulator (TI/MI) interface in
Bi$_2$Se$_3$/MnSe(111) system as an example. We demonstrate that a
gapped  ordinary bound state which spectrum depends on the interface
potential arises in the immediate region of the interface. The
gapped topological Dirac state also arises in the system owing to
relocation to deeper atomic layers of topological insulator. The gap
in the Dirac cone is originated from an overlapping of the
topological and ordinary interfacial states. This result being also
corroborated by the analytic model, is a key aspect of the magnetic
proximity effect mechanism in the TI/MI structures.

\end{abstract}


\maketitle

To efficiently realize the potential of three dimensional (3D)
topological insulators (TIs) in spin electronic and magnetic storage
applications, it is advisable to integrate the TI layers into hybrid
heterostructures containing the layers of ferromagnetic (FM) or
antiferromagnetic (AFM) materials \cite{Qi,Fujita}. One has to tune
these heterostructures in such a way that the spectrum of the 3D TI
surface electron states should be easily accessible to an exchange
field influence of FM (AFM) layer without significant spin dependent
scattering of these states on magnetic ions. At the same time, the
spin dependent transport of carriers in 3D TI should be
controllable. These demands are not easily feasible because of a
number of obstacles. On the one hand, a serious problem is to find
the FM or AFM material which forms a high-quality interface with the
TI material and at the same time provides a strong magnetic
interaction with it. On the other hand, the physics of the exchange
coupling at the TI/FM (AFM) boundary is not yet well understood. In
principle, there exist different possibilities to provide an
exchange field influence from a magnetic material on the surface
electron states of a 3D TI.

One way is to use the effect of the surface magnetic order in 3D TIs
with chemisorbed magnetic impurities. While the local magnetic
moments of impurities are arranged inside a thin layer with the
thickness of the order of their diffusion length into TI and form
magnetically ordered overlayer, the region of spin polarization of
carriers near the 3D TI surface may be significantly larger due to
magnetic proximity effect. This type of order can be realized by
deposition of magnetic ions of 3d-metals (Mn, Fe, Cr, Co) on
Bi$_2$Te$_3$, Bi$_2$Se$_3$ or Sb$_2$Te$_3$
\cite{Wray,Scholz,Honolka,Ye}. When the surface concentration of
ions is relatively high an indirect exchange coupling among their
local magnetic moments mediated by the surface states of 3D TI can
arise \cite{Henk}. As a result the system becomes unstable with
respect to FM order with a spontaneous magnetization along the
normal axis, which is accompanied by opening a gap in a spectrum of
the Dirac surface states \cite{Henk,Tanaka,Rosenberg,Men,Li}.
However, if the energy of exchange coupling is smaller than the
reciprocal lifetime of the TI surface state due to impurity disorder
scattering, both the FM ordering and the energy gap should be
suppressed.

Another way to induce magnetic order on the surface of 3D TI's is
coating with an external FM or AFM overlayer. This has been done in
Ref.~\cite{Vobornik} where FM order was induced in
Bi$_{2-x}$Mn$_x$Te$_3$ ($x$=0.09) by a magnetic proximity effect
through the deposited Fe overlayer in the temperature range well
above the intrinsic Curie temperature of the bulk
Bi$_{2-x}$Mn$_x$Te$_3$. Unfortunately, in the case of metallic FM
(AFM) materials the TI surface states near the TI/FM(AFM) interface
should be significantly altered due to their hybridization with the
bulk states of  FM (AFM) metal. While DFT calculations are absent
for this type of structures, one can suppose that the spectrum of
the TI surface states should even lose helical features due to the
alternation, since the density of states of the metallic FM (AFM)
overlayer is much larger than the density of surface states of the
3D TI.

The film of traditional FM (AFM) insulator (below called magnetic
insulator, MI) adjacent to TI is a most promising candidate to
manipulate the helical states of 3D TI by means of magnetic
proximity effect \cite{Garate}. Such a way may diminish the surface
scattering via continuum states of the magnetic film, contrary to
the metallic film case. Recently, several MIs (EuO, EuS, EuSe, MnSe,
MnTe, RbMnCl$_3$) with compatible magnetic structure and relatively
good lattice matching with TIs (Bi$_2$Te$_3$, Bi$_2$Se$_3$ or
Sb$_2$Te$_3$) are identified, and the best candidate material is
found to be the large gap AFM semiconductor MnSe \cite{Luo}.

\begin{figure}
  \includegraphics[width=0.7\columnwidth]{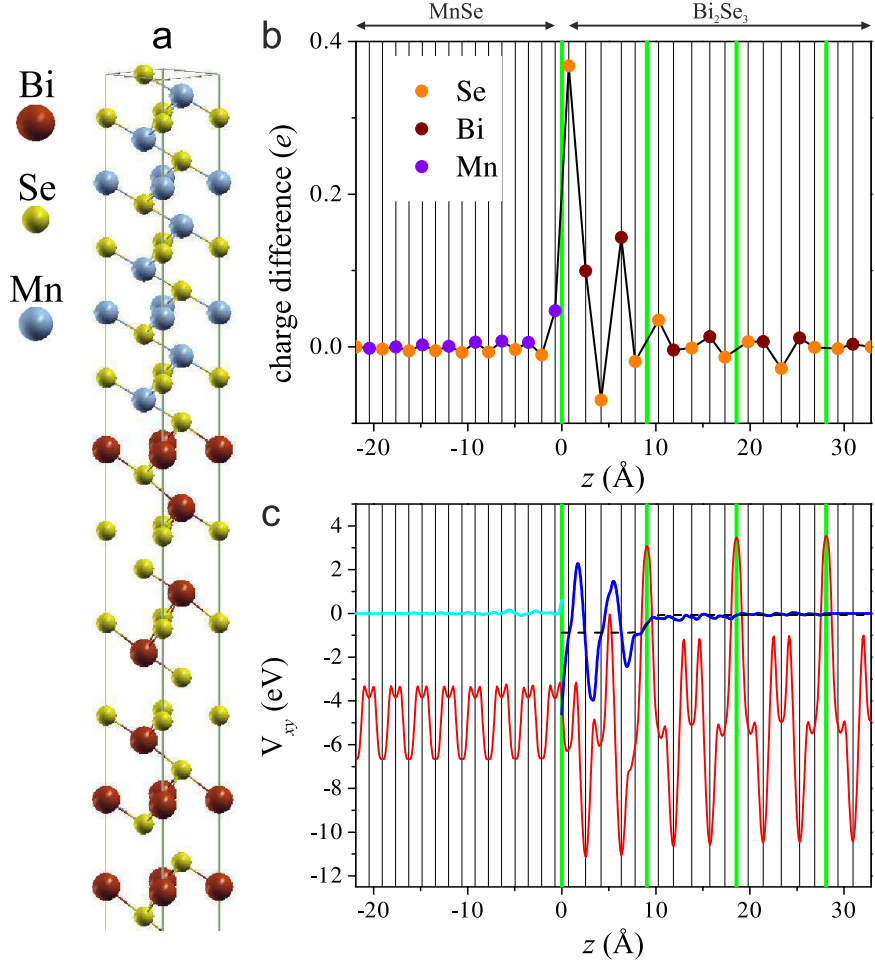}
    \caption{(Color on-line) (a) Crystal structure of the Bi$_2$Se$_3$/MnSe(111) fcc-type interface (half of the supercell is shown);
    (b) Atomic charge difference with respect to the
     electronic charge of bulk-like atoms; (c) Total electrostatic potential averaged over $xy$ planes
     $V_{xy}$ (red line) and change of the potential with respect to that in the central parts of
     the MnSe ($\Delta V_{\rm MI}$, light blue line) and Bi$_2$Se$_3$ ($\Delta V_{\rm TI}$, dark blue line)
     slabs. Dashed line shows Boltzmann fit for $\Delta V_{\rm TI}$. Vertical black lines mark position
     of atomic layers; vertical green lines denote borders of Bi$_2$Se$_3$ QLs;
     $z=0$ corresponds to the interfacial plane.}
\label{fig1}
\end{figure}

In the present paper, we study the physics of magnetic proximity
effect at the TI/MI interface within first-principles DFT
calculations for the Bi$_2$Se$_3$/MnSe(111) system as an example. We
show that two types of interfacial bound states (referred as the
topological and interfacial ordinary state, respectively) appear at
the TI side of the interface. These states have different physical
origins, spatial distributions and energy spectra. Namely, the
topological state stems from a breaking of the $\mathbb{Z}_2$
invariant of TI at the boundary with MI. This state is located
relatively distant from the interface plane; its spectrum is gapped
and lies inside the bulk energy gap of TI. In contrast, the
interfacial ordinary state results from the crystal symmetry
breaking at the TI/MI interface. This state is located nearby the
interface and is strongly spin polarized, its spectrum is gapped and
lies far below the bulk energy gap of TI due to the band bending at
the TI side of the interface.

For structural optimization and electronic bands calculations we use
the Vienna Ab Initio Simulation Package \cite{VASP1,VASP2} with
generalized gradient approximation (GGA) \cite{PBE} to the exchange
correlation potential. The interaction between the ion cores and
valence electrons was described by the projector augmented-wave
method \cite{PAW1,PAW2}. The Hamiltonian contains scalar
relativistic corrections, and the spin-orbit interaction (SOI) is
taken into account by the second variation method \cite{KH}. To
describe correctly the highly correlated Mn-$d$ electrons we include
the correlation effects within the GGA+$U$ method as developed in
Ref.~\cite{U}.

To simulate the Bi$_2$Se$_3$/MnSe(111) heterostructure, the in-plane
lattice constant of the MnSe is fixed to that of Bi$_2$Se$_3$. The
most stable structure of bulk MnSe is a cubic NaCl-type lattice,
with antiferromagnetic ordering along the [111] direction
\cite{Klosowski}. The optimization of MnSe with the fixed parameter
in (111) plane leads to 10\% contraction of Mn-Se interlayer
distance in the [111] direction. The typical values of correlation
parameters $U = 5.0$ eV, and $J = 1.0$ eV \cite{Youn,Amiri} are
appropriate for cubic MnSe, while for distorted MnSe they give Mn
$d$ states within the gap. For this reason $U = 6.0$ eV, repelling
Mn $d$ bands into the conduction band is used for the
Bi$_2$Se$_3$/MnSe(111) heterostructure calculations. It is important
that the orthorhombic distortion of MnSe keeps the Mn magnetic
moment equal to $\pm 4.58 \mu_{\rm B}$ (it is $\pm 4.57 \mu_{\rm B}$
in cubic structure with experimental lattice parameters) as well as
antiferromagnetic ordering along the [111] direction. It means that
in the direction perpendicular to the interface the orthorhombically
distorted MnSe will provide the same magnetic exchange coupling as
the cubic MnSe.

\begin{figure}
  \includegraphics[width=0.7\columnwidth]{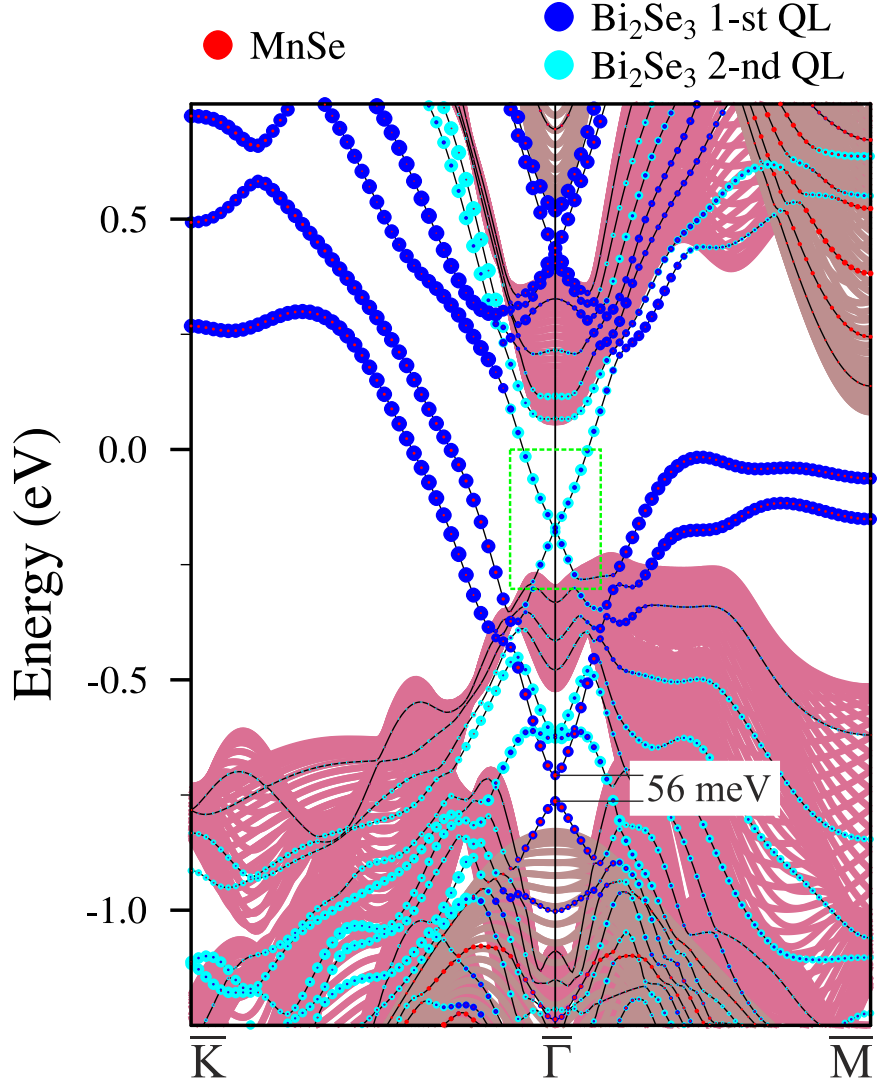}
    \caption{(Color on-line) Electronic structure of Bi$_2$Se$_3$/MnSe heterostructure.
  Size of red circles correspond to the weight of the states in
  5 layers of MnSe, closest to interface; dark and light blue circles denote weight of the states in
  the first and second QLs of Bi$_2$Se$_3$ near interface.  The projected bulk bands are shown in
  palevioletred and rosybrown for Bi$_2$Se$_3$ and MnSe, respectively.}
 \label{fig2}
\end{figure}

The Bi$_2$Se$_3$/MnSe(111) interface was studied by constructing a
superlattice composed of Bi$_2$Se$_3$ and MnSe slabs. Two types of
interfaces with Mn interfacial atomic layer are geometrically
possible: fcc-type interface, where interfacial Mn atom is situated
in the fcc-hollow position ontop of the Bi$_2$Se$_3$ slab
(Fig.~\ref{fig1}(a)) and hcp-type interface, where interfacial Mn
atom is situated in the hcp-hollow position (not shown). The
constructed supercells contain two interfaces. To avoid the
interface-interface interaction the slabs of 7 quintuple layers
(QLs) of Bi$_2$Se$_3$ and of 31(33) atomic layers of MnSe were used
for the fcc-type (hcp-type) interface. The interfacial space
(separation distance between Bi$_2$Se$_3$ and MnSe slabs) as well as
the atomic positions within the first (closest to the interface) QL
of Bi$_2$Se$_3$ and 5 near-interface atomic layers of MnSe were
optimized while interatomic distances within the middle part of both
slabs were fixed. The total energy optimization shows that the
fcc-type interface gains an energy of 115 meV with respect to the
hcp-type heterostructure. For this reason in the following we will
focus on the fcc-type Bi$_2$Se$_3$/MnSe(111) interface.

As far as the charge transfer and charge redistribution is the
common feature for any interface we have estimated this effect by
implementing the Bader charge analysis \cite{bader}. In
Fig.~\ref{fig1}(b) the atomic charge difference with respect to the
total electronic charge of central (bulk-like) atoms of Bi$_2$Se$_3$
and MnSe slabs is shown. In contrast to small oscillation of the
charge in the MnSe slab and in inner QLs of Bi$_2$Se$_3$ a large
charge redistribution is found in the interfacial QL. Such a change
in electron charge density within the first QL results in
substantial modification of the electrostatic potential which shows
modulated band-bending behavior within the interfacial QL
(Fig.~\ref{fig1}(c)). Owing to the strong modification of the 1-st
QL potential the localized states of the interfacial QL split off
from the conduction band and spread across the gap
(Fig.~\ref{fig2}). Additionally, two types of the states arise at
-0.6 -- -0.8 eV in the local bulk energy gap: the gapped (56 meV)
interfacial state and degenerate at the $\bar\Gamma$ point two
spin-split states, localized in the second QL, which are split off
from the gap edges. The latter states are similar to those which
reside near the bottom of the local valence band gap in the TIs of
Bi$_2$Se$_3$ family (they were studied in detail in Sb$_2$Te$_3$
\cite{Pauly}) but here they appear at higher energy owing to that
the lower part of this gap is filled by the MnSe bulk states. The
former, interfacial state, is similar to the state with a gap of
$\sim 54$ meV found in Ref.~\cite{Luo}, which was assumed as a
gapped Dirac cone. Next interesting feature in the spectrum is that
the topological Dirac state, being localized in the outermost QL on
the free Bi$_2$Se$_3$ surface, survives in the formation of the
interface relocating to the second QL. The small thickness of the
Bi$_2$Se$_3$ slab (4QLs) in Ref.~\cite{Luo} did not allow to catch a
veritable Dirac state, relocating into the second QL.

\begin{figure}
  \includegraphics[width=\columnwidth]{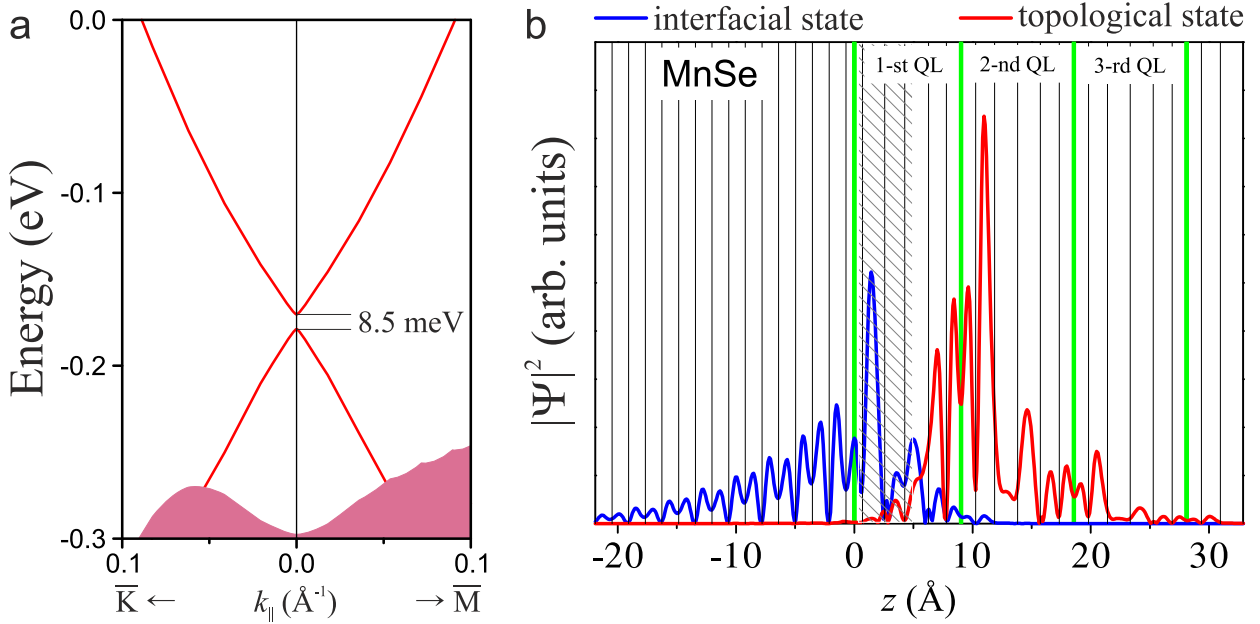}
    \caption{(Color on-line) (a) A magnified view of electronic structure of Bi$_2$Se$_3$/MnSe(111)
    [$E(k_{\|})$ ranges correspond to dashed (green) frame marked in
    Fig.~\ref{fig2}];
    (b) Spatial localization of the topological (red)
    and interfacial (blue)
    states at the $\bar\Gamma$ point, hatched area covers the Bi$_2$Se$_3$ layers with induced magnetization.}
 \label{fig3}
\end{figure}

A magnified view of the Dirac state is shown in Fig.~\ref{fig3}(a).
As one can see, the cone is gapped at the $\bar\Gamma$ point.
Fig.~\ref{fig3}(b) shows that the topological state tends to leave
the layers with induced magnetization. Thus the gap of 8.5 meV in
the Dirac cone is provided by an overlap of the topological and
spin-polarized interfacial states within the first QL
(Fig.~\ref{fig3}(b)). The induced magnetization at the interface is
limited to three layers of Bi$_2$Se$_3$. The magnetic moment on the
second layer Bi is 0.04 $\mu_{\rm B}$ while it is smaller on Se
atoms ($\leq 0.01$ $\mu_{\rm B}$).

To elucidate the obtained results we develop an analytical model
for the magnetic proximity effect in the TI/MI heterostructure,
which is based on a recently proposed method to describe the
formation of the bound in-gap electron states at the interface
between 3D TI and a normal insulator \cite{Men-13}. Our model
differs from an approach routinely used in the study of the TI/MI
interface \cite{Garate} when the existence of the exchange term in
the phenomenological Hamiltonian of the 2D Dirac-like states is
simply postulated.

We write the full electron energy of the TI/MI heterocontact in the
following form:
\begin{equation}
\Omega=\int\limits_{z>0}d{\mathbf r}\Theta^+(\mathbf r)H_t(-i\nabla)\Theta(\mathbf r)+\int\limits_{z<0}d{\mathbf r} \Phi^+(\mathbf r)H_m(-i\nabla)\Phi(\mathbf r)+\Omega_I,
 \label{Omega000}
\end{equation}
\begin{equation}
\Omega_I=\int d{\mathbf r}[\Theta^+(\mathbf r)V(\mathbf r)\Phi(\mathbf r) + \Phi^+(\mathbf r)V^+(\mathbf
r)\Theta(\mathbf r)].
 \label{OmegaIII}
\end{equation}
In the TI half-space ($z>0$), the four bands $\mathbf k \cdot
\mathbf p$ Hamiltonian $H_{t}$, proposed in Ref.~\cite{Zhang} for
the narrow-gap semiconductors of the Bi$_2$Se$_3$ family,
describes the low energy and long wavelength bulk electron states
near the $\Gamma$ point of the Brillouin zone. In the MI
half-space ($z<0$), the electron states are modeled within the
effective mass approximation by the four bands Hamiltonian without
SOI, $H_{m}$. The intrinsic magnetization of MI is assumed to be
perpendicular to the TI/MI interface plane ($z=0$). The spinors
$\Theta(\mathbf{r})$ and $\Phi(\mathbf{r})$ are smooth and
continuous envelope functions in the right and left half-spaces,
respectively. The term $\Omega_{I}$ involves intermixing of the TI
and MI electron states at the interface via the effective
potential of hybridization $V(\mathbf{r})$.

Following the variational procedure \cite{Men-13}, one can show that
the energy functional (\ref{Omega000})-(\ref{OmegaIII}) has two
distinct bound states: ordinary and topological states with the
exponentially decaying asymptotics far from the interface:
$\Phi(z\rightarrow-\infty)=0$ and $\Theta(z\rightarrow\infty)=0$.
One of them, the interfacial ordinary state, is localized near the
interface and exponentially decays into the TI half-space with a
scale length $z_o$. The charge redistribution near the interface
causes significant shift of the ordinary state spectrum relative to
the TI bulk energy spectrum \cite{Men-13}. As seen from the
Boltzmann fit for $\Delta V_{\rm TI}$ in Fig.~\ref{fig1}(c), in the
Bi$_2$Se$_3$/MnSe system, the band bending on the TI side of the
interface is $\sim$0.8 eV so that interfacial ordinary state is sunk
into the region of the bulk valence band. Furthermore, due to the
hybridization (\ref{OmegaIII}) with the orbitals of the MI outermost
layer that exhibits the out-of-plane magnetization $M$, the
interfacial ordinary state becomes spin polarized.

Another state at the interface, the topological state, is not
directly influenced by the interface potential since
$\Theta_{t}(z=0)=0$. This topological state is remote from the
interface at the distance $z_t<z_o$; note the lengths $z_{o}$ and
$z_{t}$ are determined by the material parameters of TI. Thus the
topological state experiences a magnetic effect of MI through the
magnetic proximity effect, when the exchange field induced by the
spin polarization of the interfacial ordinary state penetrates
deep into the TI half-space. The magnetic proximity-induced gap at
the Dirac point in the electron spectrum of the topological state
is estimated as $\Delta\sim JMS|V|^{2}$, where $J$ is the exchange
interaction strength in TI, $S\sim
\int_{0}^{\infty}dz|\Theta_{o}\Theta_{t}|^{2}$ is the overlap
integral of the ordinary and topological states in TI. The
analytical results in detail will be represented elsewhere.

In summary, on the base of DFT calculations performed on the
Bi$_2$Se$_3$/MnSe system, we scrutinized the magnetic proximity
effect at the 3D TI/MI interface. We have shown that the charge
redistribution and the mixing of the TI orbitals with the MI
orbitals at the interface causes drastic modifications of the
electron structure near the TI/MI interface. The calculation data
reveal the presence of the interfacial ordinary state confined
within the nearest interfacial QL of TI which slowly decays into MI.
This state is shifted downwards to the local energy gap owing to the
near-interface band bending. The state is gapped and spin polarized
due to the hybridization with the MI states. On the other hand, the
topological state is mainly localized in the second QL showing an
overlap with the interfacial ordinary state which in this way
mediates indirect exchange coupling between MI and the topological
state. The topological state acquires the energy gap at the Dirac
point proportional to the overlap of the topological and interfacial
ordinary states. The key DFT results are supported by the analytical
model. The unveiled mechanism of the magnetic proximity effect in
the TI/MI structure provides a pathway to integrate TIs in
spintronic devices.

%We acknowledge partial support from the Basque Country Government,
%Departamento de Educaci\'{o}n, Universidades e Investigaci\'{o}n
%(Grant No. IT-366-07), and the Spanish Ministerio de Ciencia e
%Innovaci\'{o}n (Grant No. FIS2010-19609-C02-00).


\begin{thebibliography}{99}

\bibitem{Qi} X.-L. Qi, R. Li, J. Zang, S.-C. Zhang, Science {\bf
323}, 1184 (2009).

\bibitem{Fujita} T. Fujita, M. B. A. Jalil, and S. G. Tan, Applied
Physics Express {\bf 4}, 094201 (2011).

\bibitem{Wray} L. A. Wray, S.-Y. Xu, Y. Xia, D. Hsieh, A. V. Fedorov,
Y. S. Hor, R. J. Cava, A. Bansil, H. Lin, and M. Z. Hasan, Nature
Physics \textbf{7}, 32 (2011).

\bibitem{Scholz}  M. R. Scholz, J. S\'{a}nchez-Barriga, D. Marchenko, A. Varykhalov,
A. Volykhov, L. V. Yashina, and O. Rader, Phys. Rev. Lett.
\textbf{108}, 256810 (2012).

\bibitem{Honolka} J. Honolka, A. A. Khajetoorians, V. Sessi, T. O. Wehling, S.
Stepanow, J.-L. Mi, B. B. Iversen, T. Schlenk, J. Wiebe, N. B.
Brookes, A. I. Lichtenstein, Ph. Hofmann, K. Kern, and R.
Wiesendanger, Phys. Rev. Lett. \textbf{108}, 256811 (2012).

\bibitem{Ye} M. Ye, S.V. Eremeev, K. Kuroda, E.E. Krasovskii, E.V. Chulkov, Y.
Takeda, Y. Saitoh, K. Okamoto, S. Y. Zhu, K. Miyamoto, M. Arita, M.
Nakatake, T. Okuda, Y. Ueda, K. Shimada, H. Namatame, M. Taniguchi,
A. Kimura, Phys. Rev. B \textbf{85}, 205317 (2012).

\bibitem{Henk} J. Henk, M. Flieger, I. V. Maznichenko, I. Mertig, A. Ernst, S.
V. Eremeev, and E. V. Chulkov, Phys. Rev. Lett. {\bf 109}, 076801
(2012).

\bibitem{Tanaka} Y. Tanaka, T.Yokoyama, and N. Nagaosa, Phys. Rev. Lett. {\bf 103}, 107002 (2009).

\bibitem{Rosenberg} G. Rosenberg and M. Franz, Phys. Rev. B {\bf85}, 195119 (2012).

\bibitem{Men} V. N. Men'shov, V. V. Tugushev, E. V. Chulkov, JETP Lett. \textbf{94}, 629 (2011).

\bibitem{Li} Z. L. Li, J. H. Yang, G. H. Chen, M.-H. Whangbo, H. J. Xiang, and X.
G. Gong, Phys. Rev. B {\bf85}, 054426 (2012).

\bibitem{Vobornik} I. Vobornik, U. Manju, J. Fujii, F. Borgatti, P. Torelli, D.
Krizmancic, Y. S. Hor, R. J. Cava, and G. Panaccione, Nano Lett.
\textbf{11}, 4079 (2011).

\bibitem{Garate} I. Garate and M. Franz, Phys. Rev. Lett. \textbf{104}, 146802 (2010).

%\bibitem{Luo}  W. Luo and X. L. Qi, arXiv:1208.4638
\bibitem{Luo}  W. Luo and X. L. Qi, Phys. Rev. B {\bf 87}, 085431 (2013).

\bibitem{VASP1} G.~Kresse, J.~Hafner, Phys. Rev. B {\bf 48}, 13115 (1993).
\bibitem{VASP2} G.~Kresse, J.~Furthm\"{u}ller, Comput. Mater. Sci. {\bf 6}, 15 (1996).
\bibitem{PBE} J.P.~Perdew, K.~Burke, M.~Ernzerhof, Phys. Rev. Lett. {\bf 77}, 3865 (1996).
\bibitem{PAW1} P.E.~Bl\"{o}chl, Phys. Rev. B {\bf 50}, 17953 (1994).
\bibitem{PAW2} G.~Kresse, D.~Joubert, Phys. Rev. B {\bf 59}, 1758 (1999).
\bibitem{KH} D.D. Koelling, B.N. Harmon, J. Phys. C {\bf 10}, 3107 (1977).
\bibitem{U} A.I. Liechtenstein, V.I. Anisimov and J. Zaanen, Phys. Rev. B {\bf
52}, R5467 (1995).

\bibitem{Klosowski} P. Klosowski, T.M. Giebultowicz, J.J. Rhyne, N. Samarth, H.
Luo, and J. Furdyna, J. Appl. Phys. {\bf 69}, 6109 (1991).

\bibitem{Youn} S.J. Youn, Journal of Magnetics {\bf 10}, 71 (2005).
\bibitem{Amiri} P. Amiri, S. J. Hashemifar, and H. Akbarzadeh, Phys. Rev. B {\bf 83},
165424 (2011).

\bibitem{bader} W. Tang, E. Sanville, and G. Henkelman, J. Phys.: Condens. Matter
{\bf 21}, 084204 (2009).

\bibitem{Pauly} C. Pauly, G. Bihlmayer, M. Liebmann, M. Grob, A. Georgi, D.
Subramaniam, M. R. Scholz, J. S\'{a}nchez-Barriga, A. Varykhalov, S.
Bl\"{u}gel, O. Rader, and M. Morgenstern, Phys. Rev. B {\bf 86},
235106 (2012).

\bibitem{Men-13} V. N. Men'shov, V. V. Tugushev, E. V. Chulkov, Pis'ma Zh. Eksp. Teor. Fiz. {\bf 97}, 297 (2013) [JETP  Lett.
\textbf{97}, (2013) (in press)].

\bibitem{Zhang} H. Zhang, C.-X. Liu, X.-L. Qi, X. Dai, Z. Fang, and S.-C. Zhang, Nat. Phys. \textbf{5}, 438 (2009).



\end{thebibliography}
\end{document}